# Investing in the Quantum Future – State of Play and Way Forward for Quantum Venture Capital

Christophe Jurczak[1]

Quantonation, Paris, France & Boston, USA

**Abstract:** Building on decades of fundamental research, new applications of Quantum Science have started to emerge in the fields of computing, sensing and networks. In the current phase of deployment, in which quantum technology is not yet in routine use but is still transitioning out of the laboratory, Venture Capital (VC) is critical. In association with public funding programs, VC supports startups born in academic institutions and has a role to play in structuring the priorities of the ecosystem, guiding it toward applications with the greatest impact on society. This paper illustrates this thesis with a case-study: the experience of the first dedicated quantum fund, Quantonation I, chronicling its impacts on the production of scientific knowledge, job creation and funding of the industry. The paper introduces concepts to support the emergence of new startups and advocates for funding of scale-up quantum companies. The paper concludes with proposals to improve the impact of the industry by taking steps to better involve society-at-large and with a call for collaboration on projects focused on the applications with a large societal benefit.

**Introduction – Venture Capital, research and innovation**

Venture Capital (VC) is a form of private equity financing that investors provide to startup companies believed to have significant long-term growth potential. Business historians trace the origins of the institutional venture capital industry to 1946, when Harvard Business School professor Georges Doriot formed the American Research & Development Corporation in Boston to invest in ventures developed during World War II: "*Doriot articulated and practiced many of the key principles of venture investment that continue to this day. These guideposts include: the intensive scrutiny (and frequent rejection) of business plans prior to financing, the provision of oversight as well as capital, the staged financing of investments, and the ultimate return of capital and profits to the outside investors that provided the original funding* [1]." It is characteristic of Venture Capital that most ventures in a fund's portfolio will fail. But, generally, a relatively small fraction succeeds at a scale that more than makes up for those that do not. This extreme ratio of failure to success is called the "power law" and is what (counter-intuitively) drives VC [2].

Today "*Venture capital is associated with some of the most high-growth and influential firms in the world. For example, among publicly traded firms worldwide, seven of the top eight firms by market capitalization in May 2020 had been backed by venture capital prior to their initial public offerings: Alphabet, Apple, Amazon, Facebook, and Microsoft in the United States, and Alibaba and Tencent in China*" [1]. VC-backed companies account for 41% of total U.S. market capitalization and 61% of U.S. R&D spending. VC drives job creation to such an extent that 41% of employees of US companies which went public in the last 45 years work for VC-backed companies, amounting to 6M jobs in 2020 [3] [4]. The rate of job growth at VC-backed companies is 8x that at non-VC-backed companies [5].

Thanks to specific U.S. government policies implemented at the end of the 1970s, this country leads the industry with 53% of all VC investments globally originating there [6] [3]. Before the 1970s, the rest of the G7 countries had a comparable rate of high-performing firm creation (in terms of their share of the top 300 companies globally), but the US has since created 2-3x as many top 300 companies per decade as the rest of the G7. Nearly all of those have been VC-backed (81/87).

Nevertheless, many argue that VC financing has limitations in its ability to advance technological change [1] and question whether other models for the funding of technological innovation might not be more adequate, for example Research Impact Bonds (RIBs) that support public-private partnerships [7] or Focused Research Organizations (FROs) for long term moonshots [8].

---

[1] Contact: christophe@quantonation.com

I will argue here that VC funding of technologies originating in Quantum Science is a very efficient mode for the financing of the "Laboratory to Startup" stage, drawing upon lessons learnt from the first Quantum-focused early-stage Venture Capital Fund (VCF) Quantonation I. In our experience, having in an investment management team in addition to traditional VC skills, commercial business experience and a deep understanding of both the science itself and the dynamics within the academic communities generating the science, is paramount for identifying valuable investment opportunities and helping them succeed, without being distracted by misplaced hype [9].

I will share in this paper some lessons-learnt that can help to guide action toward creating and scaling more successful startups to build technologies that the world will benefit from.

1. Funding quantum technologies – public and private investment

"Deep-Tech" startups look to commercialize cutting-edge science and technology and disrupt incumbent technologies; they now constitute 20% of VC funding. A good definition of Deep-Tech is *"technology that was science fiction in the past but is reality today; it pushes the boundaries of human capability through novel research and directed commercialization"* [10]. Also from a recent report on Deep-Tech investing: *"forward-looking investors understand that deep tech offers attractive rewards because its companies tackle large problems"* [11]. Deep-Tech startups are different from more familiar "Tech" startups founded in the categories of internet, mobile, e-commerce, collaboration, messaging sectors, and many more. Deep-Tech startups are usually spin-offs from research laboratories founded by scientists, but – while attractive for investors – generally face scientific, technological and market risks that necessitate greater time to reach maturity and establish product-market fit.

Quantum Technologies fall into this Deep-Tech category. The industry receives funding from three primary sources: governments, corporate investment and VC investment. Funding has increased drastically since the "Hello World" date for Quantum Technologies when IBM gave access to a few superconducting qubits through the cloud [12]. VC investors have poured funding into startups in this sector at an accelerated pace since 2020 (Figure 1(a)) with more than $2Bn invested in 2021 and 2022, and an estimated $1.5Bn invested in 2023. The drop-off in 2023 reflects a general drop-off in VC investments rather than one particular to Quantum Technologies. These amounts are still modest relative to other sectors, including both AI [13] and Blockchain at its peak [14], with tens of $Bn invested by the VC industry per year.

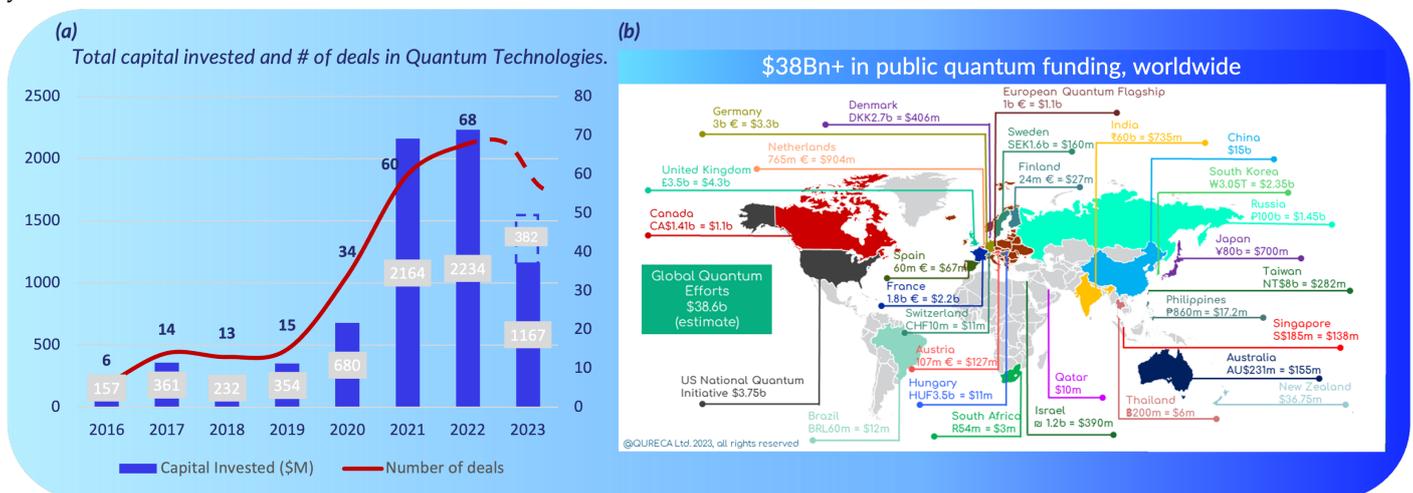

Figure 1 (a) Total capital invested by Venture Capital firms (including Initial Public Offerings) in Quantum Technologies and number of deals executed (sources: Quantonation, BCG, Hello Tomorrow, McKinsey & company, The Quantum Insider) (b) Public funding for Quantum Technologies worldwide, as of July 2023 [15].

Simultaneously, public funding toward Quantum Technologies has scaled considerably, with governments around the world deploying budgets through structured initiatives (*e.g.* in the US [16] and France [17]). An estimated $38Bn [18] has been committed to endpoints as varied as fundamental research, applied research, infrastructure and the public funding of startups and corporations through grants. Finally, internal funding of Quantum Technologies by large corporates has also grown considerably to an estimated $8Bn (over the last 5 years) [19].

It is this three-way mix of funding sources that is fueling progress in Quantum Science toward "Economic Quantum Advantage" [20], *i.e.* deployment of quantum technologies in devices and application software for day-to-day applications.

## 2. Quantonation I – Lessons learnt from the first quantum-focused VC fund 2018-2023

*2.1. Formation of Quantonation I and impact indicators*

Quantonation I (QI) was the first Venture Capital Fund created by management firm Quantonation Ventures [21], and the first quantum-focused VCF in the industry. It was structured as a French FPCI (*Fonds Professionnel de Capital Investissement*), with successive closings in 2021 and 2022, reaching a total amount of capital raised among public and private Limited Partners (LPs) of 91 M€. It was preceded in 2018-2020 by a financial holding that raised 5.2 M€, with assets merged into the main fund at first closing.

Focused on investment in Quantum and Deep Physics startups at early stages, from the laboratory "(pre-seed", in VC phase terminology) to early commercial development generating substantial revenues (Series B), QI's application focus ranges from computing to networks and sensors, combining hardware and software developments, investment in full-stack companies from qubits to quantum computing applications, as well as enabling technologies and "less-quantum" but similarly oriented projects (Deep Physics). Quantonation I's portfolio of 25 startups is located in Europe and North-America. First Quantonation investments have been performed at the incorporation of the companies, alongside the founders, for 1/3 of the portfolio.

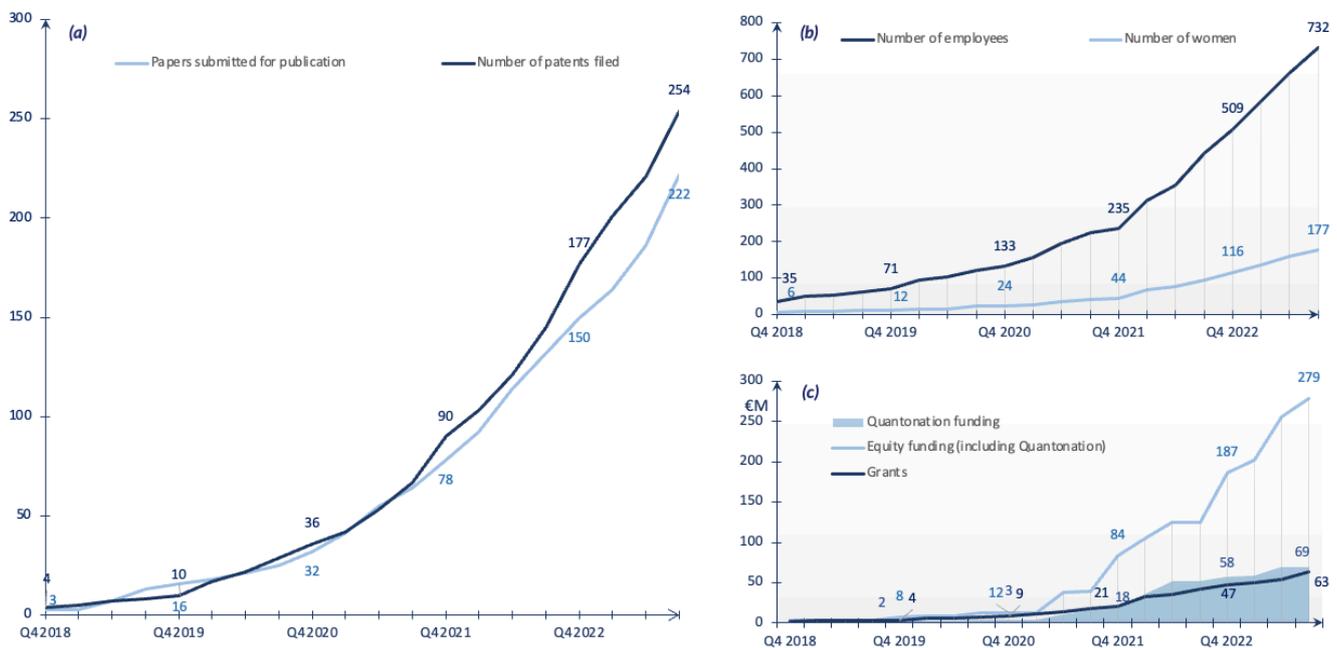

*Figure 2 – Impact Indicators for the Quantonation I VC fund, out of a portfolio of 25 startups as of 11/1/2023 on 69.3 M€ deployed. (a) Scientific papers submitted for publication in peer-reviewed journals, and number of patents filed. (b) Number of jobs created. (c) Equity investment (including Quantonation) and grants (public funding).*

As illustrated in Figure 2, QI has delivered impact on each criterion tracked since its inception in Q4 2018 (including the indicators for the financial holding mentioned above): total funding raised, jobs created (more than 730 over the period, of which only 25% are women), numbers of papers submitted and patents filed. While the fund itself can't obviously take credit for all the value created, it has undeniably had an impact as an initiator that is significant. Other criteria are also tracked, fulfilling the requirement of the Principles for Responsible Investment [22] Quantonation Ventures is a signatory of. The six PRI principles offer a menu of actions for incorporating Environmental, Social and Governance (ESG) issues into investment practice and are increasingly incorporated into the by-laws of VCFs, as well as thorough reporting requirements to LPs who want to qualify and quantify the impact of their investment.

*2.2. Scientific publications by quantum startups: why it matters and specificities*

Quantum startups in the QI portfolio publish scientific work in two primary forms, scientific papers in peer-reviewed scientific journals and scientific whitepapers. Figure 2 (a) represents submissions to peer-reviewed journals, with a publication time typically between 6 and 24 months after submission. The generation of new scientific knowledge shared with the broader scientific community, in coordination with patenting of Intellectual Property, is important for raising the profile and prestige of a startup, establishing its leadership in the industry, attracting talent and validating its approach through peer-review. Publishing is also the best approach to mitigate "quantum hype" which – if left unaddressed – is likely to disrupt the priorities, structure and organization of fundamental scientific research [9].

This class of publication is written up independently or in collaboration with the university that a startup has spun out from. A few high-profile examples of papers from spinouts in the QI portfolio include: PASQAL and Multiverse, working on early applications of analog quantum computing in finance [23], Qubit Pharmaceuticals, working on the simulation of chemical reactions using quantum processors [24], Nord Quantique, working on hardware-efficient approaches to error-correction [25], Qunnect, working on entangled photon sources for quantum networks [26], Qnami, working on metrology of magnetic memories [27] and QphoX, working on microwave-to-photonics transduction [28].

Beyond scientific journals, startups would do well to direct more of their attention toward publishing whitepapers that place a company's technology in the broader context of the industry or present a roadmap toward long term applications of a company's technologies (for example, the potential of neutral atoms for quantum computing [29] or rare-earth ions for networking [30]). This class of publication is of particular interest to investors in the "due diligence" process while assessing the promise of a new technology.

Another important ambition for startup publications moving forward is to benchmark their results against industry standards. Assessing "quantum advantage" [31] or "quantum utility" [32] in comparison to the abilities of incumbent technologies – whether for computing or sensing, for example – is key to motivate VC funding for a company. This benchmarking is difficult for early-stage companies because of a lack of proper human and financial resources to assess the landscape, but open-source approaches such as MetriQ [33] look set to change that and are likely to become widespread. The newest generation of scientific papers, such as PASQAL's recent study on quantum-enhanced graph neural networks [34], are tackling the challenge by making systematic comparisons with the most recent alternatives and should be emulated. Metrics and benchmarks provide a common language for evaluating quantum computers [35], and quantum technologies in general.

Finally, the majority of publications by quantum startups are in the applied sciences. These are generally written by entities with short histories and are often multi-disciplinary (*e.g.* at the intersections of computer sciences and physics or quantum science and computational chemistry). This makes publication in traditional scientific journals challenging. It would be desirable for the academic community to embrace alternative kinds of scientific publication – with the same rigorous peer-reviewing processes – but that are more open to multidisciplinary work and work from startups. Early examples are the new community journal Quantum [36] as well as PRX Energy [37] and PRX Life [38].

*2.3. The key role of public funding*

As illustrated in Figure 2 (c), QI's intervention in the industry has crowded-in 4.95x its original investments in the form of investments by co-investors (other VCFs, or public funds *e.g.* BpiFrance or European Innovation Council) and public grants (for example Sensorium's funding by TII Abu Dhabi [39] on an application with a high sustainability impact). This suggests that the early-stage Venture Capital ecosystem is in rude health, with many funds interested in co-investing and leading investment rounds.

Grants – which are often made available in the context of national strategies (see Figure 1 (b)) – are an essential component of the funding strategies of Deep-Tech startups, in particular in the early stages (first 2-3 years) when they match or exceed VC funding. It is crucial for pre-seed startups to leverage shared infrastructure to reduce the amount of capital they need to raise and also develop as much as possible the technology and company concept *before* incorporating. As soon as VC funding is allocated, typical Deep-Tech VCFs have a time horizon of 8-12 years depending on when they are investing in their own fundraising and deployment cycle, which might end up being too short for some projects. Choosing the proper VC investor in view of its funding allocation is key for a startup, and assessing the capital needs (*aka* business plan) properly over time is fundamental. A venture studio supported by shared infrastructure (see 3.1) is a way to fund "pre-mature" ideas.

Figure 2 (c) doesn't account for a very important part of public funding, which is more indirect but already substantial and critical for the future of the quantum industry: public procurement. In previous technology revolutions, the role of public procurement has been fundamental. As an illustration, in 1962 the U.S. government bought 100% of the integrated circuit production of the industry through NASA and MIT as part of the Apollo program [40]. That figure declined to 85% in 1963, 85% in 1964 and 72% in 1965. In Europe, the EuroHPC institution and associated EU member states have been very active in purchasing Quantum Processing Units (QPUs) for their public supercomputing centers [41] and these orders are already a substantial part of the revenue streams of startups. This is an approach that should be extended to networking, communications and sensing.

## 3. Recommendations for the Future

The development of quantum technologies in the last decade has been astounding. However, to be sustainable, the industry needs to produce a steady flow of new startups, support the emergence of VCFs for growth stages, and pioneer new models of collaboration to drive research into these technologies' most impactful applications.

*3.1. Lab. to Startups & the emergence of Quantum Venture Studios*

Most of QI's portfolio of startups were spun out from universities or co-founded by academics. Figure 3 illustrates that after the initial wave of quantum startups (in the U.S. and then in Europe, peaking in 2018) the industry is now witnessing a slow-down in company creation (although new regions are emerging, in particular in the Asia-Pacific). This is to be expected for an emerging industry and many factors have driven this trend. This means that the community should work better toward facilitating the creation of entrepreneurial projects, some funded by VCs while others will thrive better with different funding models (self-financing, debt, crowdfunding, *etc.*). But the one thing it certainly does not represent is a lack of progress, funding or ideas.

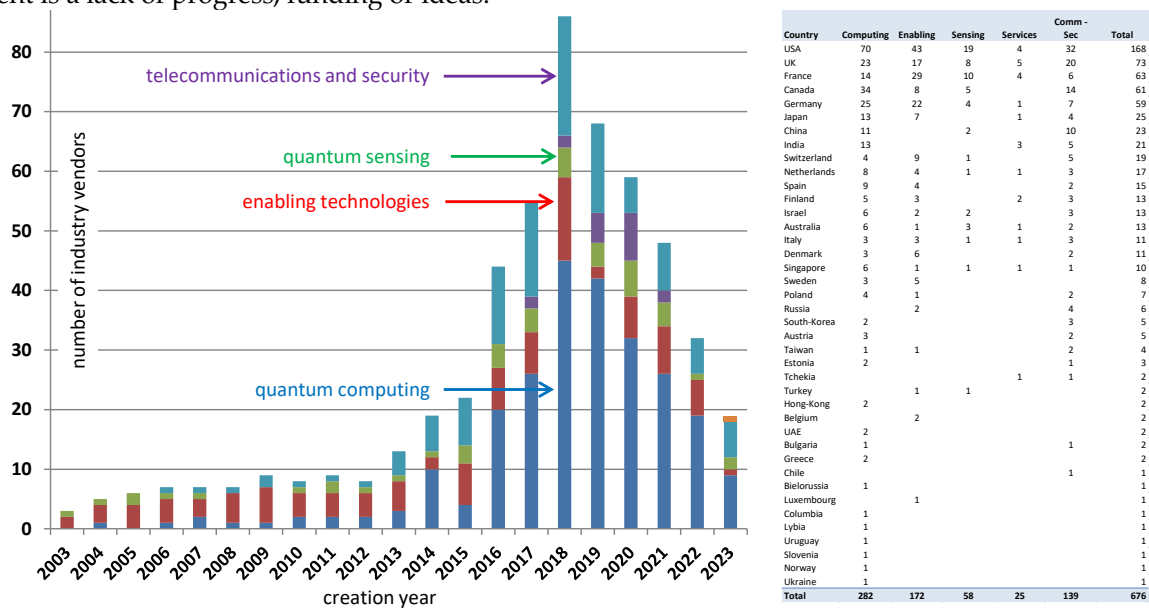

*Figure 3 – Rate of startup creation in Quantum Technologies and geographic distribution [19]*

A first approach is to help scientists turn their research into successful startups. There are several excellent programs that do this *e.g.* Deep-Tech Founders in France [42] and Activate in the US [43]. The specificities of VC funding are not always properly understood by scientific entrepreneurs and bad decisions, in terms of repartition of the share capital between founders and academic advisors or unbalanced Intellectual Property rights, might make the company un-investable by VCs. The Association of Dutch Universities recently published a common standard for deal-terms with academic spin offs [44] and such approaches should be encouraged.

The notion of "innovation ecosystems" has emerged as a promising approach in the context of joint value creation between various stakeholders [45]. An emerging actor, specifically dedicated to facilitating the creation of startups, is the *Venture Studio*. Backed by Quantonation, the first one in the world is a public-private partnership based in the DistriQ ecosystem in Sherbrooke, Canada [46] [47]. Structures like this aim to co-create companies with aspiring founders, associating entrepreneurs in residence and coaches before company creation and leveraging existing research infrastructure in universities and national laboratories, such as dilution fridges, lasers, micro/nanofabrication facilities, on a pay-per-use model, radically reducing capital expenditures in the first stages. Such a venture studio is by its nature local. That allows it to build on regional ecosystems and regional or national public funding. I anticipate and encourage the establishment of several similar initiatives worldwide, ideally working on the same principles but with modes of funding adapted to local ecosystems.

Last, these efforts should not be done at the detriment of funding of *fundamental* research. Getting familiar with disruptive new science takes time; "practicing mRNA" took 30 years and led ultimately to the emergence of BioNTech and a leadership position in mRNA vaccines. Similarly Nobel Laureate Alain Aspect made it abundantly clear at the 1st Alain Aspect Symposium on the applications of Quantum Science [48] that it took more than 40 years from his groundbreaking research implementing Bell's inequalities to the commercialization of devices based on the properties of quantum entanglement . "Practicing Quantum" is an essential part of the journey toward the creation of successful startups, and these are born in ecosystems where fundamental and applied Quantum Science is flourishing.

*3.2. Funding of Scale-Up Quantum companies*

While Quantonation, with QI and soon with a successor fund Quantonation II, is funding the early stages in the life of startups, alongside like-minded VCFs, other categories of funds and instruments are needed to finance the scale-up phase (typically after the "Series C" stage in VC terminology), once the scientific and engineering risks are tackled, with a focus on making products more reliable, affordable and available globally [11]. These rounds of funding need typically more and sometimes much more than $100m for Quantum Computing startups, and above $30m per round for enabling technologies, sensing and networking companies. The need for such funding becomes crucial now that many companies created in 2017-2020 (see Figure 3) are reaching maturity.

Public equity markets can offer a substantial pool of capital for scale-ups (*e.g.* IonQ), less dilutive financing such as hybrid capital (quasi-equity or venture debt) could be an important growth-enabler, but there is in any case a need for late-stage lead VC investors to get comfortable with quantum technologies, the horizons and the exit options, and several institutions, *e.g.* in the European Union [49], have identified the opportunity at hand for dedicated funding instruments.

Many Quantum startups are generating revenues already, in the Quantum Computing and Networking markets through customer projects and public procurement, and for commercially ready applications in sensing and enabling / supply-chain hardware and software. There are opportunities for Merger & Acquisitions and exits that should materialize in the near future. This will be a signal of maturity for the whole quantum industry ecosystem.

*3.3. New modes of collaboration toward tackling societal challenges with Quantum Technologies*

The number of quantum startups is on the rise. As they reach technical maturity and move toward commercial deployment, it is worth exploring how academic institutions, large corporates and startups are currently working together and how they can work together better in the future.

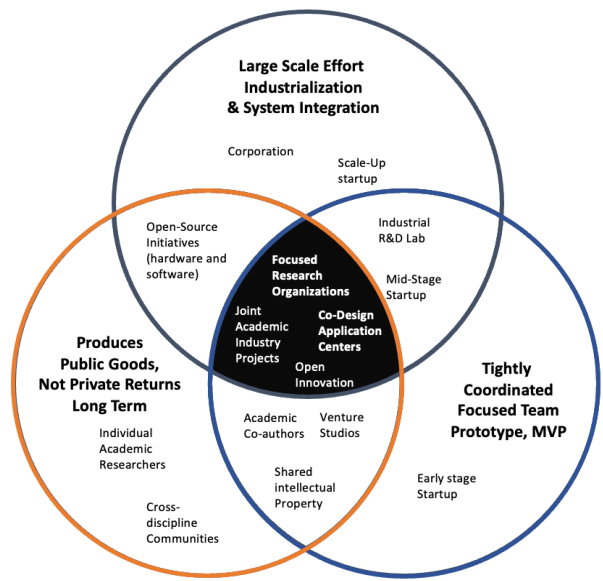

*Figure 4 – The landscape of Science and Innovation and new collaboration models (inspired from [8])*

One approach to structuring future interactions is road mapping exercises. Road mapping exercises have lead, for example, to the EU Quantum Flagship [50] and the NSF roadmap on Quantum Networks [51]. An example of a cross-discipline collaborative project is the Quantum Energy Initiative [52], developing approaches to estimate and minimize the resource costs of quantum technologies. The Q4Climate initiative [53] led to a recent conference with participation from academia, government agencies, large corporates and startups from various disciplines with the willingness to collaborate [54]. GESDA has announced in 2023 the launch of the Open Quantum Institute [55], a new "Quantum for All" initiative focused on the UN Sustainable Development Goals. The non-profit Unitary Fund [55] is one of the most visible initiatives in the development of open source hardware [56] and software [57]. And there are many academy-industry-startup collaborative projects on specific subjects in the field.

Nonetheless, what is still lacking is a collaborative approach to finding impactful applications of quantum technologies to "*demonstrate the value of quantum computers for societal challenges*" and for "*specialists [to] work together to create narratives around the usefulness of quantum technologies*" [58]. Over the last five years, startups and corporates have generated many proof-of-concept projects [59] before technologies were mature enough to demonstrate a Quantum Advantage (QA), but have not been able to generate a clear assessment of what the "winning applications" would be, especially for Quantum Computing. We are reaching now – with several technologies – ranges of parameters that put QA on the short-term horizon and that calls for new approaches. Two suggestions are given below.

**Create and expand centers for co-design of applications**. Co-design is an approach to design that involves all stakeholders in the design process of products. One example in the quantum industry is the Brookhaven Co-design Center for Quantum Advantage [60], which connects stakeholders with workstreams on the challenges associated with Quantum Computing. I am proposing similar initiatives with a more application-focused approach, leveraging hybrid quantum-classical technologies and open-source technologies. These Co-design Quantum Application Centers (CQAC) should focus on societal challenges – the Energy Transition, Aging and Health, Climate Change Mitigation, Food Security ... – building on clear science-led assessments about the potential of quantum technologies [61] [62]. They would include metrics and robust benchmarks, including ESG criteria and Life-Cycle Analysis, updated with progress of incumbent technologies provided by end-users and leverage shared infrastructure and promote collaborations between like-minded initiatives worldwide. They would generate application-oriented White Papers for the community to build actionable road maps. They would be funded by public-private partnerships, making the best of the resources provided by governments and VCs for participating startups (Figure 1) but could also leverage new models and philanthropy such as FROs [8]. Application-specific Venture Studios could be associated with these Centers.

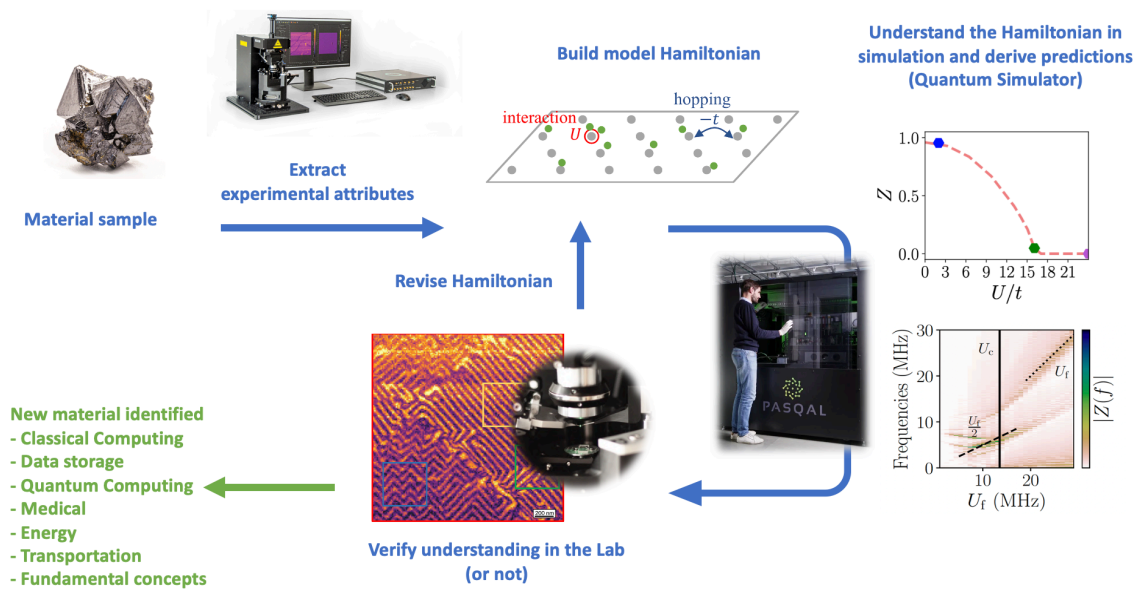

*Figure 5 – Quantum-first quantum materials design loop for new applications (pictures courtesy of Qnami and PASQAL, inspired from [63])*

**Create a co-design center focused on quantum matter and materials.** One intriguing aspect of this approach would be to look for applications where we know QA exists already [64]. This includes the field of Quantum Materials, where Analog Quantum Computers – also known as Quantum Simulators – running on several hundred qubits are perfectly operational in academic laboratories and through cloud services provided by startups such as PASQAL. Quantum Sensors, made of Nitrogen Vacancy centers in diamond, such as provided by Qnami, are ideal tools to explore these materials [65] and new Atomic Layer Deposition technologies are providing new synthesis capabilities [66] in addition to Pulsed Laser Deposition, Molecular-Beam Epitaxy and sputtering. Quantum Matter has been a hot topic of academic interest for several decades now [67] [68] but applications have always been approached as an afterthought. Combining Quantum Technologies that are *for sure* providing a competitive advantage should lead to applicative breakthroughs for Quantum Matter, the computational-sensing loop in Figure 5 is hopefully a template for a new approach in that direction. Let's note that, through quantum materials discovery, today's first generation of Quantum Computers would potentially be used to design future and more impacting ones.

## 4. Conclusions

Quoting Nobel Laureate Professor Frank Wilczek from his 2016 paper "Physics in 100 years", "*Announcements of the end of physics are decidedly premature, as are closely related proclamations of post-empirical physics. We can and will make progress on many fronts. We can and will gain new insights into and control over concrete, real-world physical phenomena. Brilliant prospects lie ahead*" [69]. This was written before IBM's "Hello World" moment, and I am convinced brilliant prospects lie ahead for Quantum Technologies.

At this early stage in the construction of the global quantum innovation ecosystem, investors, philanthropists, startup founders, academics, corporate managers and government representatives should think collectively about the design of the ecosystem they want to build and acknowledge how innovation and industrialization are different from fundamental research. VC has a role to play in this construction, not only through financial contributions. There is a path for Quantum Technologies not to always be / just be "technologies of the future" but to offer less ephemeral futures than past ones in the terms of Philosopher of Science Vincent Bontems [70]. The Historian of Science Jürgen Renn examines in his book "The Evolution of Knowledge" [71] the role of knowledge in global transformations going back to the dawn of civilization, while providing perspectives on the complex challenges confronting us today in the Anthropocene—this new geological epoch shaped by humankind. Let's use the unique occasion we have here to witness the birth of a new knowledge-based industry, to apply new principles and be more ambitious.

**Acknowledgements**

Thanks to Jean-Gabriel Boinot-Tramoni and Nicolas Fleury for collecting data and designing Figure 1 and Figure 2. Thanks to Peter Chapman for bringing my attention to NASA's impact on the semiconductor industry in the 60s, to Matt Langione for suggesting VC statistics, to Christian Sarra-Bournet for pointing to the GESDA initiative, to Zachary Yerushalmi for the mRNA / quantum "practicing" analogy and to Olivier Ezratty. And huge thanks to the team at Quantonation and many amazing founders at Quantonation portfolio companies for comments and suggestions.

9 of 11